\documentclass[osajnl,twocolumn,showpacs,amsmath,amssymb,10pt]{revtex4-1} 

\usepackage{graphicx}
\usepackage[english]{babel}
\usepackage{textcomp}

\begin{document}

\title{Optimization of integrated polarization filters}
\author{Denis Gagnon}
\author{Joey Dumont}
\author{Jean-Luc D\'eziel}
\author{Louis J. Dub\'e}
\email{Corresponding author: ljd@phy.ulaval.ca}

\affiliation{D\'epartement de physique, de g\'enie physique et d'optique \\Facult\'e des Sciences et de G\'enie, Universit\'e Laval, Qu\'ebec G1V 0A6, Canada}

\begin{abstract}
This study reports on the design of small footprint, integrated polarization filters based on engineered photonic lattices. Using a rods-in-air lattice as a basis for a TE filter and a holes-in-slab lattice for the analogous TM filter, we are able to maximize the degree of polarization of the output beams up to 98 \% with a transmission efficiency greater than 75 \%. The proposed designs allow not only for logical polarization filtering, but can also be tailored to output an arbitrary transverse beam profile. The lattice configurations are found using a recently proposed parallel tabu search algorithm for combinatorial optimization problems in integrated photonics.
\end{abstract}


\maketitle

Polarization is a physical dimension of light that can be exploited to increase the rate of transmission of information in optical communications \cite{Winzer2014}. For instance, polarization beam-splitters based on modal birefringence in integrated waveguides may enable transmission rates up to 400 Gbps in optical networks \cite{Perez-Galacho2014}. Integrated polarization manipulation is also critical to accelerating electrons using dielectric structures \cite{Breuer2013}. These examples are but a small sample of important applications that have moved the design of integrated elements dedicated to polarization management to the forefront of photonics research. Some existing solutions for tailoring the polarization of light at the microscale level include subwavelength gratings \cite{Chien2004, Nguyen-Huu2011}, chains of coupled optical microspheres \cite{Darafsheh2013}, Raman processes \cite{Xiong2013}, metasurfaces \cite{Avayu2014} and photonic crystals \cite{Pottier2006, DeLaRue2012}. In parallel to these developments, various photonic crystal (PhC) inspired devices have also been proposed, such as near-field beam shapers \cite{Vukovic2010, Gagnon2012, Gagnon2013}, lenses \cite{Sanchis2004, Marques-Hueso2013}, waveguide bends \cite{Xing2005} and waveguide couplers \cite{Andonegui2013}. The design process of these nanophotonic devices is almost always based on the optimization of a primitive PhC lattice -- or grid of scatterers -- using \textit{metaheuristics}, optimization algorithms based on empirical rules for exploring large solution spaces \cite{Talbi2009}.

The aim of this Letter is to optimize small footprint integrated devices combining two functionalities: beam shaping and polarization filtering. We show that two basic scatterer grids can be used for this purpose, namely a rods-in-air (RIA) lattice for TE polarization filtering and a holes-in-slab (HIS) lattice for TM polarization filtering. This choice is motivated by the band structure of each basic photonic lattice. Moreover, the polarization filters proposed are experimentally feasible as RIA lattices can be fabricated routinely using electron beam lithography of amorphous silicon films \cite{Xu2010}, and HIS lattices can be produced by etching inclusions in a high refractive index membrane of semiconductor material \cite{DeLaRue2012}. 

Although polarization selective beam-splitters based on PhC bandgaps have been demonstrated in the past \cite{Pottier2006}, our designs allow not only for logical polarization filtering, but are also specifically tailored to preserve the beam shape or to transform it to specification at the device output. Our approach consists in using a metaheuristic algorithm, \textit{parallel tabu search} (PTS), to optimize a basic photonic lattice in order to achieve the required functionality \cite{Gagnon2013}, a polarized beam with a definite shape at the device output. This will be demonstrated by the generation of quasi-Gaussian polarized beams at the near-field of the device.

\begin{figure}
\centering
\includegraphics{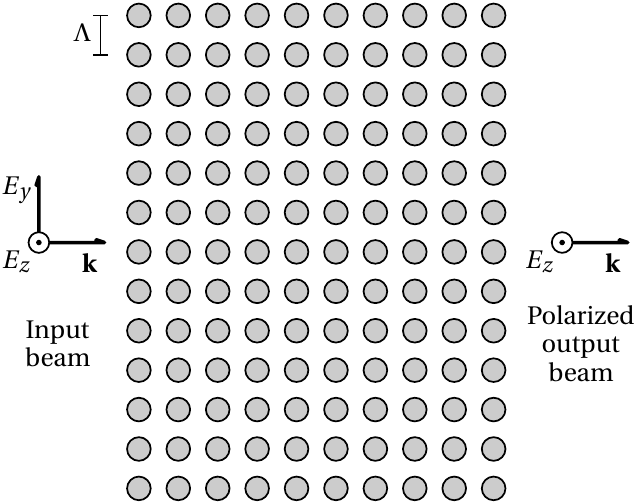}
\caption{Schematic representation of the basic photonic lattice configuration and the polarization optimization problem. In this example, the $E_y$ component of the incident beam is filtered out, resulting in a TM polarized beam. For a TE polarized beam, replace ($E_z, E_y$) by ($H_z, H_y$). The final optimized design will consist of occupied or empty scattering sites.}\label{fig4:geometry}
\end{figure}

Before proceeding with the optimization problem, it is critical to choose an adequate configuration space, in other words a basic photonic lattice. In this Letter, we use a $10 \times 13$ lattice as shown in Fig. \ref{fig4:geometry}. This geometry defines two privileged directions, namely the direction parallel to the scatterers' axis (the $z$-axis) and the beam propagation axis (the $x$-axis). The configuration space is specified by the fact that we only allow individual scattering sites to be occupied or empty in the final design, resulting in $2^{70}$ possible solutions taking a mirror symmetry into account. This basic lattice geometry was successfully used in a previous optimization study, for the conversion of a Gaussian beam to coherent Hermite-Gauss type beams \cite{Gagnon2013}.

\begin{figure}
\centering
\includegraphics{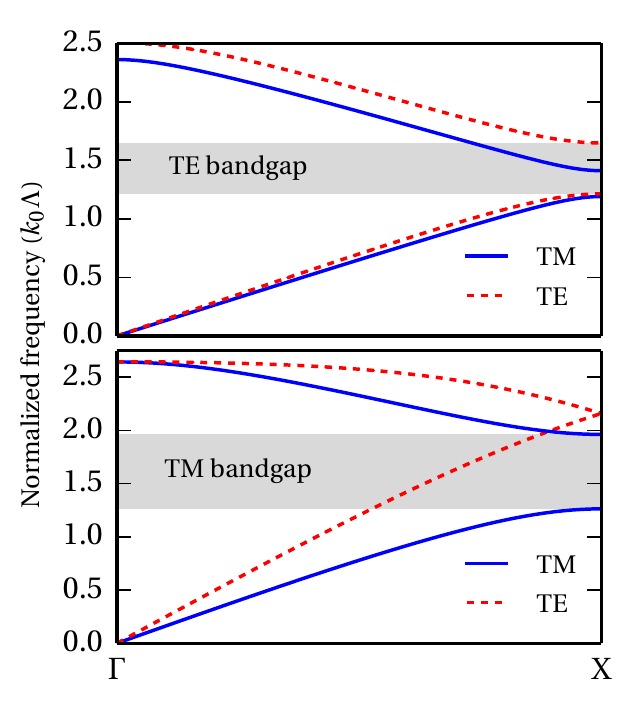}
\caption{Band structure for two different square lattices of cylinders (diameters $D=0.6\Lambda$). (Top) HIS configuration with refractive index of slab $n=2.76$. (Bottom) RIA configuration with rods of refractive index $n=3.3$ embedded in air. Eigenmodes were computed using the \textsc {MIT Photonic bands} software package \cite{Johnson2001}.}\label{fig4:bands}
\end{figure}

To obtain a TM polarized output beam ($E_y = 0$), we choose the HIS design, whereas to obtain a TE polarized beam we adopt the RIA design. These choices are motivated by the band structure of both photonic lattices. As can be seen from Fig. \ref{fig4:bands}, the HIS lattice exhibits a wider directional bandgap for the TE polarization in the $\Gamma - \mathrm{X}$ direction. This implies that the TE component of the beam ($E_y$) is more strongly scattered, making the HIS lattice suitable for filtering this polarization out and favoring the TM polarization. 
For the device to operate near that bandgap, the diameter of all air holes is set to $D = 0.6 \Lambda$, where $\Lambda$ is the lattice constant. We use an effective refractive index $n = 2.76$, corresponding to a thin silicon slab at $\lambda \sim 1.5$ \textmu m \cite{Chutinan2000}. In contrast, we choose a RIA lattice for the TE polarizer with the rods refractive index set to $n=3.3$. Similarly, this lattice exhibits a bandgap for the TM polarization, meaning that it strongly scatters the TM component of the beam ($H_y$).
The incident beam wavenumber is set to $k_0 = 1.76 / \Lambda$, with a half-width $w_0=2.5\Lambda$. Such a beam could in principle be generated using an integrated waveguide \cite{Xu2010}. 
The value of $k_0$ is chosen to fall near the bandgap of both HIS and RIA lattices. 
Although the Bloch modes expansion yielding the band diagrams does not strictly hold for our final optimized configurations (as they are neither periodic nor infinite), this approach provides a useful design tool for polarization filters. 
Moreover, integrated polarization selective beam splitters based on PhCs have been reported to exhibit an effective bandgap despite only three rows of scatterers being present in the final design \cite{Pottier2006, DeLaRue2012}.

Once the solution space has been defined, the next step is to formulate the optimization objectives. The problem consists in finding a lattice configuration which, when illuminated with a Gaussian input beam, produces a polarized beam that also matches a specific Gaussian profile in a given target plane, although in principle both the input and output profiles can be arbitrary \cite{Gagnon2012, Gagnon2013}. In two dimensions, this beam shaping problem can be formulated as the minimization of the following objective function \cite{Dickey2005, Gagnon2012}
\begin{equation}\label{eq:g1}
g = \dfrac{\int \big||u_z(x_0,y)|^2 - |\bar{u}_z(x_0,y)|^2 \big| dy}{\int |\bar{u}_z(x_0,y)|^2 dy } 
\end{equation}
where $x_0$ is the location of the target plane, $u_z(x_0,y)$ is the computed EM field on the target plane (either $E_z$ for a TM polarizer or $H_z$ for a TE polarizer) and $\bar{u}_z(x_0,y)$ is the required beam profile at the device output. For a given configuration, the resulting beam profile $u_z(x_0,y)$ can be computed using a two-dimensional generalized Lorenz-Mie theory (2D-GLMT). This computation method -- the speed of which is crucial to the optimization procedure -- is detailed in refs. \cite{Elsherbeni1992, Nojima2005, Gagnon2012}.

To obtain a polarized output beam, another objective function related to the degree of polarization $\mathcal{P}$ of the output beam must be optimized. We use the following definition \cite{Al-Qasimi2007}
\begin{equation}
\mathcal{P} = \dfrac{\int \langle S_x (x_0,y) \rangle_{z} dy}{\int \langle S_x (x_0,y) \rangle_{z} dy + \int \langle S_x (x_0,y) \rangle_{y} dy } = \dfrac{P_z}{P_\mathrm{tot}},
\end{equation}
where $\langle S_x \rangle_{z,y}$ is the $x$ component of the time-averaged Poynting vector, i. e. the power transmitted through the target plane $(x=x_0)$. The ($z,y$) subscripts represent the contribution of each orthogonal polarization to the Poynting vector. 
The ratio $\mathcal{P}$ is therefore equal to the power carried by the polarized portion of the beam $P_z$ divided by the total power contained in both polarizations $P_\mathrm{tot}$. To obtain a perfectly TM (TE) polarized beam, the contribution of the $E_z$ $(H_z)$ component must be maximized at the target plane (maximum possible value of $\mathcal{P}=1$). Since the device geometry does not mix polarizations, we suppose an equal incident power in both orthogonal field components, and attempt to maximize $P_z$. The flexibility of the 2D-GLMT approach makes the computation of the Poynting vector components a simple matter \cite{Nojima2005}. On a side note, since real structures either based on pillars or embedded in waveguides are not infinite in the $z-$direction, the modes are more accurately labeled as ``quasi-TM'' or ``quasi-TE'' because polarization mixing, however small, can indeed take place \cite{DeLaRue2012}. For the purpose of this work, we shall consider that no mixing occurs.

Another way of characterizing the polarization filters is by means of the ratio between the total power transmitted by the $u_z$ component of the field and the total power transmitted by the $u_y$ component. This ratio is computed from the Poynting vector components in the following way
\begin{equation}
R = \dfrac{\int \langle S_x (x_0,y) \rangle_{z} dy}{\int \langle S_x (x_0,y) \rangle_{y} dy } = \dfrac{P_z}{P_y}.
\end{equation}
Finally, another quantity of interest is the power transmission efficiency $\eta$, simply defined as the ratio between the power incident in the $u_z$ component on the polarization filter and the output power in the same field component \cite{Gagnon2012} 
\begin{equation}
\eta = \dfrac{\int \langle S_x (x_0,y) \rangle_{z} dy }{\int \langle S_x (x_{\mathrm{in}},y)  \rangle_{z} dy },
\end{equation}
where $x_{\mathrm{in}}$ is the location of the input plane. 
Both $R$ and $\eta$ are useful indicators to measure the performance of the final optimized configurations.

To sum up, the optimization problem consists in minimizing the objective function $g$ while simultaneously maximizing the degree of polarization $\mathcal{P}$ over a 70 dimensional binary search space. 
This set of objectives constitutes a combinatorial \emph{multiobjective optimization problem}. These problems are often tackled using \textit{metaheuristics}, general optimization techniques which aim to provide well conditioned solutions in a reasonable amount of time \cite{Talbi2009}. Metaheuristics are sometimes called global optimization algorithms.
Notable instances in photonics design include genetic algorithms (GAs) \cite{Sanchis2004, Vukovic2010, Gagnon2012, Marques-Hueso2013, Gagnon2013}, differential evolution \cite{Lavelle2010} and harmony search \cite{Andonegui2013}.
In a recent contribution, we have proposed the use of an alternative metaheuristic for combinatorial multiobjective optimization problems in photonics called \emph{parallel tabu search} (PTS) \cite{Gagnon2013}. 
The main feature of tabu search is that it uses an adaptive memory to escape from local minima in the solution space \cite{Glover1997, Talbi2009}. Besides, it involves fewer adjustable parameters and relies less on stochastic operators than the more commonly used GA. The net benefit is to increase the convergence speed for the sort of optimization problems considered here \cite{Gagnon2013}.

Using PTS, we have performed the optimization of the objective functions $(g,\mathcal{P})$ in order to find lattice configurations suited for polarization filtering, i.e. the conversion of an non-polarized Gaussian beam to a polarized one. Since both objectives are not independent, the solution to this multiobjective problem is not a single configuration, but rather a set of compromises between the two objectives, the Pareto set of the problem \cite{Talbi2009, Gagnon2013}.
For demonstrative purposes, a Gaussian beam with a half-width $w_0=2.5\Lambda$ is required at the device output. However, the beam shaping procedure just described could allow for the generation of arbitrary shaped polarized beams. To obtain polarization filters exhibiting high profile accuracy and high transmission efficiency, we have only retained the Pareto solutions with $g \leq 0.05$, that is an error on the output beam profile inferior to 5 \%.

\begin{figure}
\centering
\includegraphics{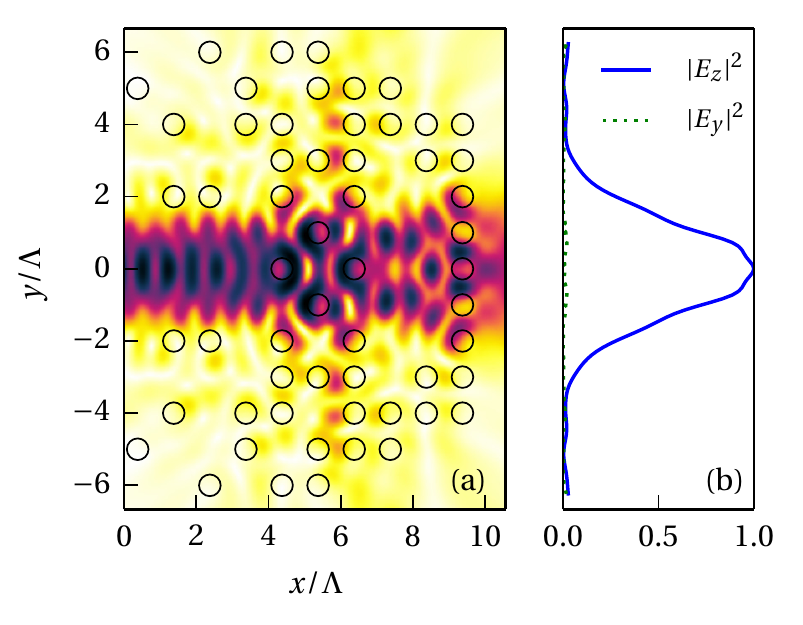}
\caption{Generation of a TM polarized Gaussian beam. (a) Optimized HIS configuration (57 scatterers) and $|E_z|$ field profile (arbitrary units). The target plane coincides with the upper limit of the $x$ axis. (b) Comparison of orthogonal polarization components along target plane. This solution is characterized by $\mathcal{P} = 0.983$, $R=59.8, g=0.044, \eta=0.759$.}\label{fig4:TM}
\end{figure}

\begin{figure}
\centering
\includegraphics{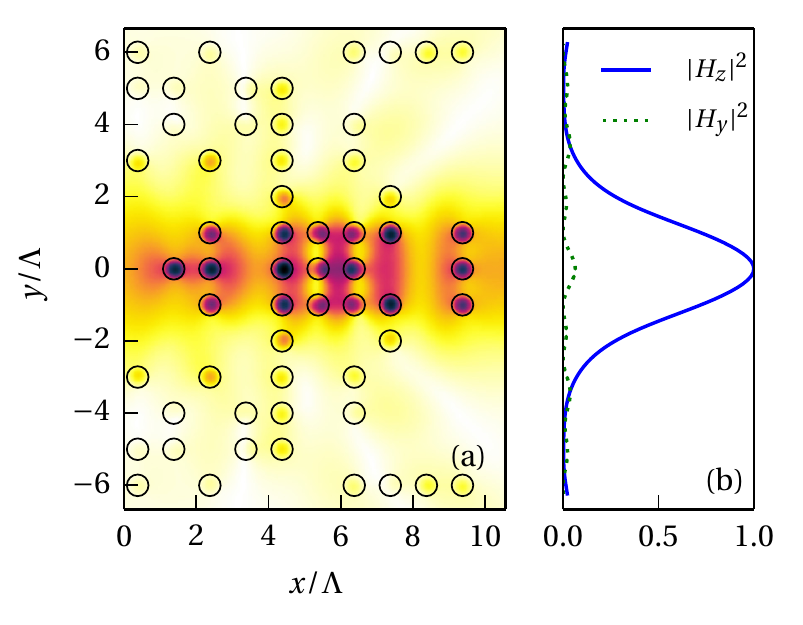}
\caption{Generation of a TE polarized Gaussian beam. (a) Optimized RIA configuration (58 scatterers) and $|H_z|$ field profile (arbitrary units). The target plane coincides with the upper limit of the $x$ axis. (b) Comparison of orthogonal polarization components along target plane. This solution is characterized by $\mathcal{P} = 0.980$, $R=48.0, g=0.028, \eta=0.890$.}\label{fig4:TE}
\end{figure}

The two best lattice configurations found (in terms of $\mathcal{P}$) satisfying this condition are shown in Figs. \ref{fig4:TM} and \ref{fig4:TE}. In both cases we are able to maximize the degree of polarization to values exceeding $\mathcal{P} = 0.980$. Alternatively, both configurations are characterized by $R \geq 48$, which means that the transmission of the preferred field component is at least 48 times higher than the filtered out component. Additionally the near-field beam shapes deviate from a Gaussian amplitude profile by less than 4.4 \% and both configurations exhibit power transmission efficiencies above $\eta = 0.75$. For the TM polarization, we also found a configuration (not shown) characterized by $\eta = 0.81$, but in that case the output beam is slightly less polarized ($\mathcal{P} = 0.978$ and $R \simeq 43.9$). We have also performed optimization using triangular primitive lattices, but we found that this procedure resulted in lower values of $\mathcal{P}$. This may be related to the fact that square grids allows for nearly complete rows to be present in the design (see Figs. \ref{fig4:TM}a and \ref{fig4:TE}a), allowing the effective bandgap effect described earlier to take place.

In summary, we have proposed small footprint integrated designs allowing for simultaneous polarization filtering and amplitude beam shaping. The designs are based on two-dimensional photonic lattices exhibiting partial bandgaps, which facilitates the filtering behavior. Using an optimization procedure based on the tabu search algorithm, we are able to maximize the average degree of polarization of the output beam up to 98 \% with a transmission efficiency over 75 \% for the TM polarizer and 80 \% for the TE polarizer. While the designs we presented allow for the generation of a Gaussian amplitude profile at the device near-field, the optimization procedure can be used for the generation of arbitrary shape beams, as shown in previous studies \cite{Vukovic2010, Gagnon2012, Gagnon2013}.

Future work includes the application of the algorithm to different beam shapes as well as a generalization to three-dimensional lattices, thereby allowing for an integrated solution to generate, for instance, radially polarized beams. Noteworthy is the fact that we were not able to obtain TE polarized beams using the HIS design and TM beams using the RIA configuration. This confirms the usefulness of the bandgap analysis as a design guide. Nevertheless, using high-order bands (where there is no bandgap) to generate polarized beams may be possible. In fact, interesting effects in high-order transmission bands of PhCs, e.g. lensing, have recently been observed \cite{Maigyte2013}. This could also be beneficial from an experimental standpoint, as the resulting polarization filters could accommodate wider beams.

The authors acknowledge financial support from the Natural Sciences and Engineering Research Council of Canada (NSERC) and computational resources from Calcul Qu\'ebec. J.D. and J.L.D. are grateful for a research fellowship from the Canada Excellence Research Chair in Photonic Innovations.

%

\end{document}